**TITLE PAGE** <u>Original</u>

(3) Title:

Generic phase diagram of "electron-doped" T' cuprates

(4) Author's Names

Osamu Matsumoto[a], Aya Utsuki[a], Akio Tsukada[b], Hideki Yamamoto[c], Takaaki Manabe[d], *Michio Naito[a]

(5) Addresses

[a] Department of Applied Physics, Tokyo University of Agriculture and Technology

Naka-cho 2-24-16, Koganei, Tokyo 184-8588, Japan

[b] Geballe Laboratory for Advanced Materials, Stanford University,

Stanford, California 94305, USA

[c] NTT Basic Research Laboratories, NTT Corporation, 3-1 Morinosato-Wakamiya,

Atsugi, Kanagawa 243-0198, Japan

[d] National Institute of Advanced Industrial Science and Technology (AIST)

Higashi 1-1-1, Tsukuba, Ibaraki 305-8565, Japan

*) Corresponding author.  Address: Department of Applied Physics, Tokyo University of Agriculture and Technology, Naka-cho 2-24-16, Koganei, Tokyo 184-8588, Japan. Tel. +81 42 388 7229; fax: +81 42 385 6255.  E-mail address: minaito@cc.tuat.ac.jp.

# Generic phase diagram of "electron-doped" T' cuprates

Osamu Matsumoto[a], Aya Utsuki[a], Akio Tsukada[b], Hideki Yamamoto[c], Takaaki Manabe[d], and *Michio Naito[a]

[a] Department of Applied Physics, Tokyo University of Agriculture and Technology

Naka-cho 2-24-16, Koganei, Tokyo 184-8588, Japan

[b] Geballe Laboratory for Advanced Materials, Stanford University,

Stanford, California 94305, USA

[c] NTT Basic Research Laboratories, NTT Corporation, 3-1 Morinosato-Wakamiya,

Atsugi, Kanagawa 243-0198, Japan

[d] National Institute of Advanced Industrial Science and Technology (AIST)

Higashi 1-1-1, Tsukuba, Ibaraki 305-8565, Japan

**Abstract**

We investigated the generic phase diagram of the electron doped superconductor, $Nd_{2-x}Ce_xCuO_4$, using films prepared by metal organic decomposition. After careful oxygen reduction treatment to remove interstitial $O_{ap}$ atoms, we found that the $T_c$ increases monotonically from 24 K to 29 K with decreasing $x$ from 0.15 to 0.00, demonstrating a quite different phase diagram from the previous bulk one. The implication of our results is discussed on the basis of tremendous influence of $O_{ap}$ "impurities" on superconductivity and also magnetism in T' cuprates. Then we conclude that our result represents the generic phase diagram for oxygen-stoichiometric $Nd_{2-x}Ce_xCuO_4$.

*) Corresponding author. Address: Department of Applied Physics, Tokyo University of Agriculture and Technology, Naka-cho 2-24-16, Koganei, Tokyo 184-8588, Japan. Tel. +81 42 388 7229; fax: +81 42 385 6255. E-mail address: minaito@cc.tuat.ac.jp.

**Introduction**

The electronic phase diagram of cuprate superconductors is a key ingredient to understand the still unresolved mechanism of high-temperature superconductivity. A particular interesting question is the differences and similarities between the hole- and electron-doped sides. The phase diagram of hole-doped high-$T_c$ cuprates shows a well-known 'dome' shape with maximal superconductivity at a doping level of about 0.15 [1]. For electron-doped high-$T_c$ cuprates, the currently accepted phase diagram is similar to the hole-doped one, except for the much narrower superconducting window [2]. This implies that electron-hole symmetry roughly holds in high-$T_c$ superconductors [3]. Based on this symmetry, the doped-Mott-insulator scenario for high-$T_c$ superconductivity has been developed [4]. However, it has to be emphasized that the above electron-hole symmetry is derived from comparing the hole and electron-doping phase diagrams in different structures, namely hole doping in the $K_2NiF_4$ (abbreviated as T) structure and electron doping in the $Nd_2CuO_4$ (T') structure.

The phase diagram of the hole-doped side [1], *e.g.* T-$La_{2-x}Sr_xCuO_{4+\delta}$, has been well established whereas that of the electron-doped side [5, 6], *e.g.* T'-$RE_{2-x}Ce_xCuO_{4+\delta}$ ($RE$ = La, Pr, Nd, Sm, Eu, and Gd) is controversial. This is because the superconductivity in the T' cuprates deteriorates seriously by the presence of impurity oxygen ($O_{ap}$) atoms [7]. Hence, in principle, it requires complete removal of $O_{ap}$ atoms to unveil the generic phase diagram of the *T'*-cuprates. Early works performed almost 2 decades ago have not paid a serious attention to this $O_{ap}$ problem, and reported a very narrow superconducting window ($x$ = 0.14 to 0.18). Later, in 1995, Brinkmann *et al*. reported that the superconducting window of $Pr_{2-x}Ce_xCuO_4$ extends down to $x$ = 0.04 via a novel reduction route [8]. However, their result has not been reproduced successfully by any other group, and its importance has been largely forgotten with time. Very recently we have succeeded in achieving superconductivity in undoped cuprates T'-$RE_2CuO_4$ ($RE$ = Pr, Nd, Sm, Eu, Gd) with $T_c$ ~ 33 K by employing a thin film synthesis technique, metal organic decomposition (MOD) [9]. The key recipe to achieve superconductivity is low-$P_{O2}$ firing, which minimizes the amount of impurity $O_{ap}$ atoms during crystallization. This discovery promoted us to reexamine the phase diagram of the doped $T'$ cuprates with impurity $O_{ap}$ atoms cleaned up by the same synthetic process as for undoped T'-$RE_2CuO_4$. In this article, we report the experimental results so far obtained for "electron-doped" $Nd_{2-x}Ce_xCuO_4$ (NCCO). Our result is quite different from the past one, showing that $T_c$ increases with $x$ from 0.15 (solubility limit of Ce in MOD) to 0.00.

**Experimental**

Superconducting NCCO thin films were prepared by MOD, using Nd naphthenate, Cu naphthenate and Ce 2-ethylhexanoate solutions. The stoichiometric mixture of solutions was spin-coated on substrates, either $SrTiO_3$ (STO) (100) or $DyScO_3$ (DSO) (110) substrates [10]. In this article, we present the results only with DSO, which gives a wider Ce solubility range than STO. The coated films were first calcined at 400 $^o$C in air to obtain precursors, then fired at 850 $^o$C ~ 875 $^o$C in a tubular furnace under a mixture of $O_2$ and $N_2$, controlling the oxygen partial pressure from 4 x $10^{-5}$ atm to 2 x $10^{-3}$ atm. Finally the films were reduced in vacuum at various temperatures ($T_{red}$) from 420$^o$C to 500$^o$C for removal of $O_{ap}$ atoms. The $T_{red}$ was optimized at each Ce concentration ($x$) so as for films to show best superconducting properties. The resultant $T_{red}$ decreased slightly (by ~ 20 $^o$C) from $x$ = 0.00 to $x$ = 0.15.

For comparison, films with no reduction were also prepared and named as "as-grown". The film thickness was typically 800 Å. The lattice constants were determined from positions of peaks in $\theta$-$2\theta$ scans in X-ray diffraction.

**Results**

Figure 1 shows the XRD patterns of reduced NCCO films with different Ce concentrations ($x$ = 0 - 0.175). The films for $x$ = 0.00 - 0.15 are single phase T' and all peaks can be indexed as (00$l$) reflection lines of T', indicating single-crystalline films achieved by solid-state epitaxy. But the film of $x$ = 0.175 showed weak and broad diffraction peaks of T', and also the existence of $CeO_2$ as a secondary phase. Furthermore the $c$-axis lattice parameter ($c_0$) of the film with $x$ = 0.175 deviates from the Vegard's law as shown below in Fig. 2, indicating that the actual Ce concentration in this film is lower than the nominal one due to precipitation of $CeO_2$. Therefore the Ce solubility limit in NCCO films lies in between $x$ = 0.15 and 0.175 with our MOD synthesis, which is much lower than $x$ ~ 0.25 with our previous MBE growth [11].

Figure 2 shows the variation of $c_0$ as a function of $x$ for reduced and as-grown NCCO films. In this figure, the bulk values reported are also plotted by a dotted line [12]. The $c_0$ values of the as-grown films roughly coincide with the bulk values but those of the reduced films are noticeably smaller, especially for lower $x$, than the bulk values. The difference is as large as 0.05 Å at $x$ = 0.00 and is reduced to 0.01 Å at $x$ = 0.15. This behavior can be understood with the following established trends: (1) $Nd_{2-x}Ce_xCuO_{4+\delta}$ is capable of containing more $O_{ap}$ atoms for lower $x$ ($\delta_{max}$ ~ 0.10-0.13 for $x$ = 0.00 and $\delta_{max}$ ~ 0.03-0.05 for $x$ = 0.15) [13-15] and (2) the $c_0$ increases almost linearly with the amount of residual $O_{ap}$ atoms [16]. Hence we think that the $c_0$ values of reduced films are representative of oxygen-stoichiometric $Nd_{2-x}Ce_xCuO_{4+\delta}$ with $\delta \sim$ 0.00, and those of as-grown films and bulk samples are expanded by the presence of a fair amount of impurity $O_{ap}$ atoms ($\delta > 0.00$).

Now we turn to the transport and superconducting properties of the MOD films. The upper panel of Figs. 3 shows the temperature dependences of resistivity of NCCO films with different $x$ and the lower panel shows a plot of resistivity at 300 K [$\rho$(300 K)] and just above $T_c^{onset}$ [$\rho(T_c^+)$] as a function of $x$. All films with $x = 0.00$ to $0.15$ showed a full superconducting transition whereas the film with $x = 0.175$ did not show zero resistance. The room temperature resistivity stays almost constant (600-800 $\mu\Omega$cm) for $x = 0.00$ to $0.15$, and it goes up for $x = 0.175$ because of the secondary $CeO_2$ phase. This indicates that there is no fundamental change in the electronic states from $x = 0.00$ to $x = 0.15$, suggesting a band-metal picture (only shifting $E_F$) for the effect of doping in NCCO.

Figure 4 shows a plot of $T_c$ of the MOD films as a function of $x$, which is compared with the one derived from past bulk works [2] and the other from our MBE works [11]. The highest $T_c^{onset}$ of the MOD films is 29 K observed in undoped $Nd_2CuO_4$ [17]. The $T_c^{onset}$ monotonically decreases from 29 K to 24 K with increasing $x$ from 0.00 to 0.15. The $T_c^{onset}$ of 24 K at $x = 0.15$ agrees with the established bulk $T_c$. We could not determine $T_c$ for $x > 0.15$ because of the solubility limit of Ce. In this region, however, the data from our MBE growth are complementary. The phase diagram derived from MBE films shows a dome shape extending from $x = 0.10$ to $0.20$. Although the $T_c$ of MBE films is always higher than the bulk $T_c$, it is lower than the $T_c$ of MOD films for $x < 0.15$. The reason is, almost certainly, insufficient $O_{ap}$ removal in MBE films. With no exceptions, MBE films with $x < 0.15$ show a low-temperature upturn in resistivity (most likely Kondo effect [7]), which is one indication of impurity $O_{ap}$ atoms remaining. This low-temperature upturn disappears for $x > 0.15$, hence we take the $T_c$ of MBE films as a good measure (or a lower bound) of intrinsic $T_c$ for $x > 0.15$. With the MOD and MBE data together, the $T_c$ of NCCO films monotonically decreases from 29 K at $x = 0.00$ and disappears around $x = 0.20$. We believe that our result, although quite different from the previous bulk one, represents the phase diagram for oxygen-stoichiometric NCCO.

**Discussion**

We discuss the implication of our results. In the previous phase diagram of T' cuprates, the superconducting region is narrow and directly adjacent to the antiferromagnetic (AF) region. Coming from the "underdoped" side, superconductivity suddenly appears at the superconducting-antiferromagnetic (SC-AF) boundary with almost maximum $T_c$ [18], suggesting the competition between SC and AF orders. One *might* imagine that the Neel temperature ($T_N$) of the AF order in the T'-$(RE,Ce)_2CuO_{4+\delta}$ *would* depend primarily on the Ce concentration but it has to be emphasized that the $T_N$ depends more dramatically on the oxygen reduction, namely the amount of residual $O_{ap}$. For a good example, AF correlations develop below $T_N \sim 150$ K in as-grown non-superconducting $Pr_{1.85}Ce_{0.15}CuO_4$ and $Nd_{1.85}Ce_{0.15}CuO_4$ specimens, but they essentially disappear in reduced superconducting samples [19]. This implies that the AF correlations in T'-$RE_{2-x}Ce_xCuO_{4+\delta}$ is not intrinsic, but is induced by impurity $O_{ap}$ atoms [20]. We claim that this statement holds not only for $x \sim 0.15$ but also over a whole range of $x$ including $x = 0.00$. The situation in T'-$RE_{2-x}Ce_xCuO_{4+\delta}$ seems to be in a sharp contrast to that in T-$La_{2-x}Sr_xCuO_4$ showing intrinsic AF order. By suppressing "impurity-$O_{ap}$-induced" AF correlations, superconductivity can be achieved at any $x$ from 0.00 to 0.20.   This is behavior generic to T'-cuprates with no impurity $O_{ap}$.

**Summary**

We have investigated the generic phase diagram of $Nd_{2-x}Ce_xCuO_4$ by careful oxygen reduction treatment for MOD films.   The $T_c^{onset}$ monotonically decreases from 29 K to 24 K with increasing $x$ from 0.00 up to 0.15 (solubility limit of Ce in MOD). According to complementary MBE data, the $T_c$ decreases further for $x > 0.15$ and disappears around $x = 0.20$.   It is concluded with the MOD and MBE data together that $T_c$ is a monotonic decreasing function of $x$ in $x = 0.00$ to 0.20.   We believe that our result, although quite different from the previous bulk one, is representative of the generic phase diagram of $Nd_{2-x}Ce_xCuO_4$.   We also claim that the AF correlations competitive to SC in T'-$RE_{2-x}Ce_xCuO_{4+\delta}$ is not intrinsic, but induced by impurity $O_{ap}$ atoms.

**Acknowledgements**

The authors thank Dr. Y. Krockenberger and Dr. J. Shimoyama for stimulating discussions, and Dr. T. Kumagai for support and encouragement.   They also thank Crystec GmbH, Germany for developing new $RE$ScO$_3$ substrates.   The work was supported by KAKENHI B (18340098) from Japan Society for the Promotion of Science (JSPS).

Figure 1    XRD patterns of $Nd_{2-x}Ce_xCuO_4$ films with different Ce concentrations ($x = 0 - 0.175$). The films for $x = 0 - 0.15$ are single phase T' and all peaks can be indexed as (00$l$) reflections of T'. The film of $x = 0.175$ shows the existence of $CeO_2$ as an impurity phase.

Figure 2    Variation of $c$-axis lattice parameter ($c_0$) of $Nd_{2-x}Ce_xCuO_4$ films as a function of $x$. The open circles and filled circles denote reduced films and as-grown films, respectively. The dotted line joints the bulk values by Uzumaki *et al*. [12]. The $c_0$ of the single-phase reduced films are noticeably smaller than the bulk values.

Figures 3    Upper: temperature dependences of resistivity for $Nd_{2-x}Ce_xCuO_4$ films and lower: Ce-doping dependences of $\rho(300\ K)$ (filled circles) and $\rho(T_c^+)$ (open circles). The $\rho(300\ K)$ and $\rho(T_c^+)$ stay almost constant up to $x = 0.15$.

Figure 4    $T_c$-versus-$x$ of $Nd_{2-x}Ce_xCuO_4$ films prepared by MOD. For a comparison, the data of bulk samples (dotted line) [2] and MBE films (triangles) [11] are also plotted. Open and filled symbols represent $T_c^{onset}$ and $T_c^{end}$. The solid line is a guide for eye.

**Figure 1**

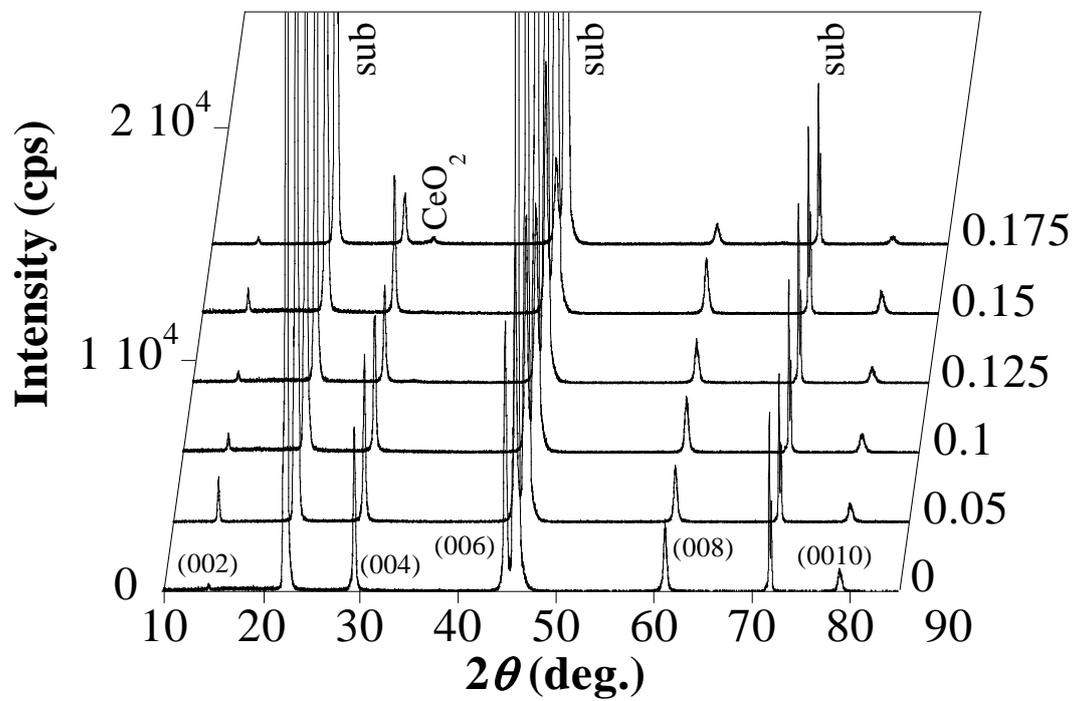

**Figure 2**

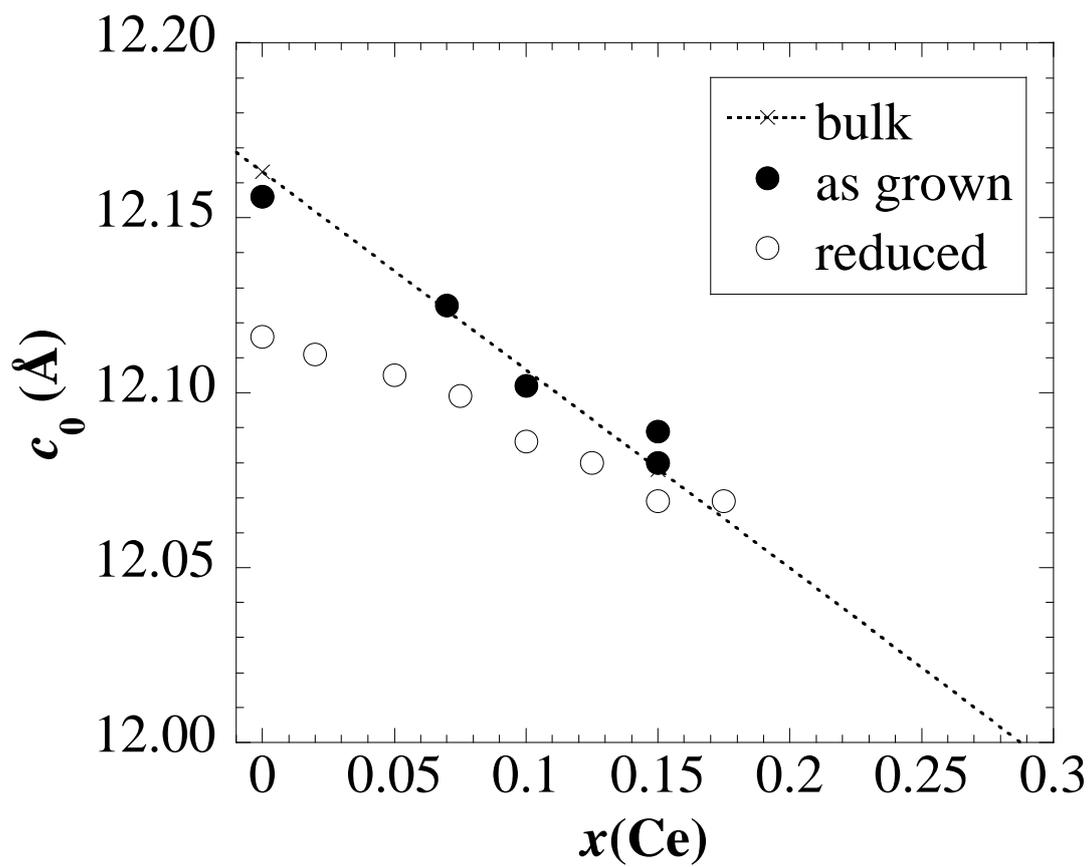

**Figure 3**

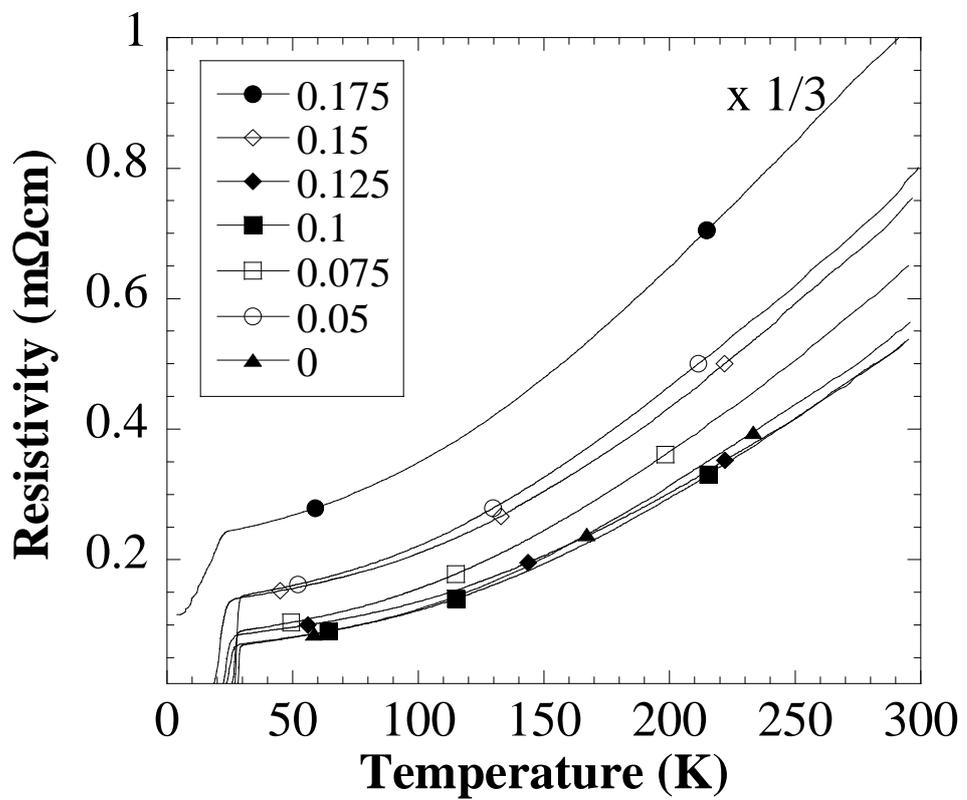

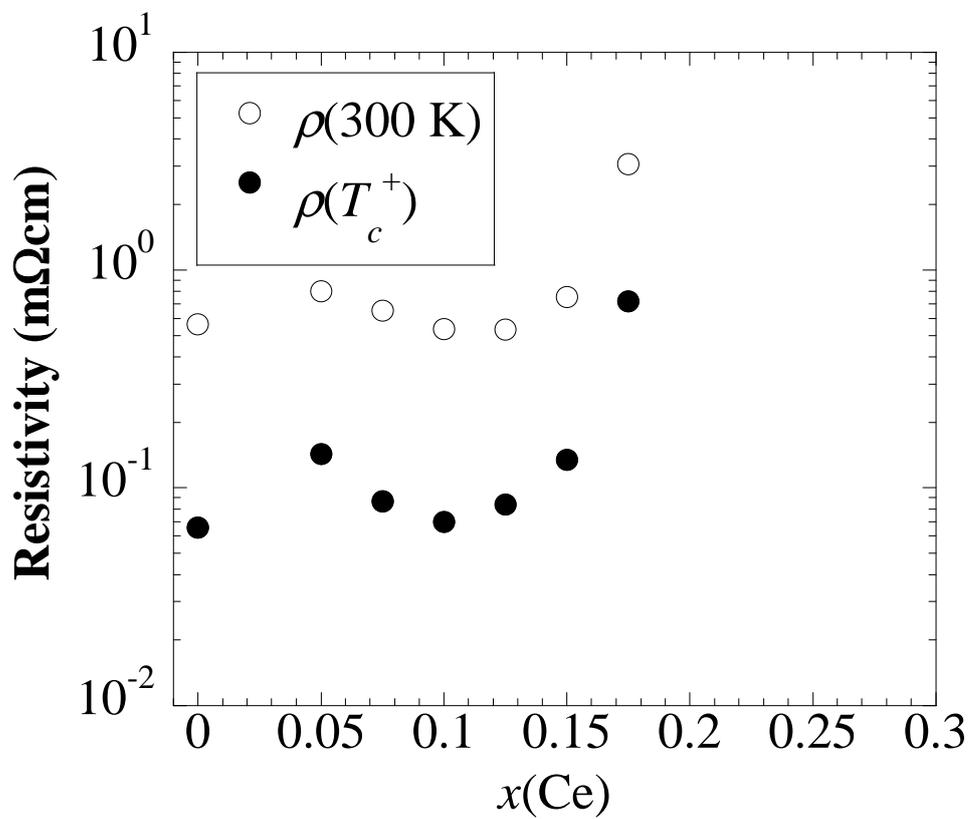

**Figure 4**

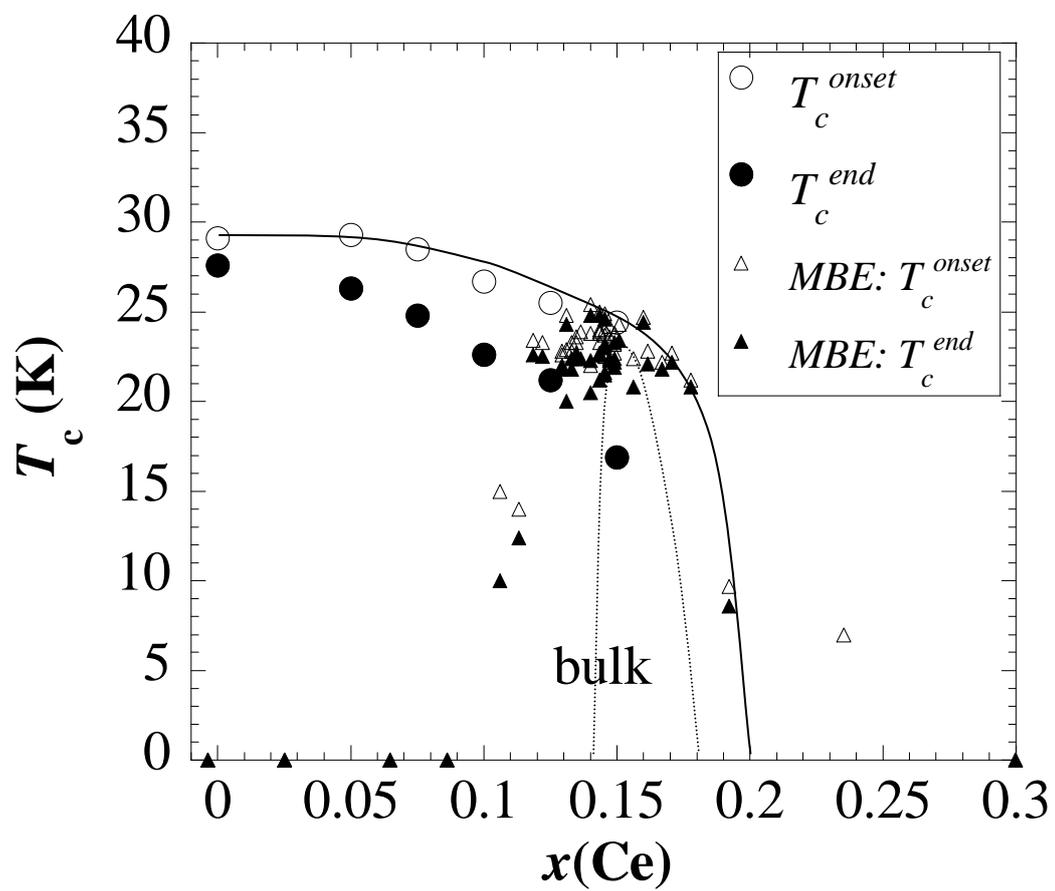